# The influence of geometry and specific electronic and nuclear energy deposition on ion-stimulated desorption from thin self-supporting membranes

Radek Holeňák[1], Michaela Malatinová[1,2], Eleni Ntemou[1], Tuan T. Tran[1], and Daniel Primetzhofer[1]

[1]*Department of Physics and Astronomy, Uppsala University, Box 516, S-751 20, Uppsala, Sweden*

[2]*Institute of Physical Engineering, Brno University of Technology, Technická 2, 616 69, Brno, Czech Republic*

**Abstract**

We investigate the dependence of the yield of positive secondary ions created upon impact of primary He, B and Ne ions on geometry and electronic and nuclear energy deposition by the projectiles. We employ pulsed beams in the medium energy regime and a large position-sensitive, time-of-flight detection system to ensure accurate quantification. As a target, we employ a single crystalline Si(100) self-supporting 50 nm thick membrane thus featuring two identical surfaces enabling simultaneous measurements in backscattering and transmission geometry. Electronic sputtering is identified as the governing mechanism for the desorption of hydrogen and molecular species found on the surfaces. Nevertheless, larger energy deposition to the nuclear subsystem by heavier projectiles as well as due to the directionality of the collision cascade appears to act in synergy with the electronic energy deposition leading to an overall increase in secondary ion yields. A higher yield of ions sputtered from the matrix is observed in transmission geometry only for B and Ne ions, consistent with the observed role of nuclear stopping.

**Introduction**

An energetic ion traversing condensed matter deposits its kinetic energy in a series of collisions with the target nuclei and electrons (Sigmund, 2017). The sum of the average energy lost in both types of interactions per unit path length is commonly defined as the stopping power

(Sigmund, 1998). The energy deposition induced by nuclear stopping, $S_n$, is based on the momentum transfer between the projectiles and recoiling target atoms which frequently induces a collision cascade. Near the surface, the subsequent emission of target constituent atoms as a result of momentum transfer is denoted as nuclear sputtering (Sigmund, 1981). In electronic stopping, $S_e$, energy is dissipated to the target by electronic excitation and projectile ionization, which also can lead to particle emission in what is commonly referred to as electronic sputtering (Assmann et al., n.d.; Gibbs et al., 1988; Johnson and Brown, 1982). Both contributions to the total stopping power, $S = S_e + S_n$, depend, for a given ion-material combination on the energy of incoming ions (Zhang et al., 2015a). Knowledge of S is straightforward mandatory for materials analysis as it can be employed to establish depth profiles at MeV and keV ion energies (Wang and Nastasi, 2009). The same knowledge is also decisive when changing the properties of materials by irradiation, e.g. controlled modifications of crystallinity, tuning implantation depth, as well as patterning 2D materials (Schmidt and Wetzig, 2013; Zhang et al., 2015b; Zhang and Weber, 2020). Finally, as the specific energy deposition in terms of stopping power and the yields of secondary particles are intimately related, techniques relying on particle emission induced by ion irradiation also demand knowledge on S.

A commonly employed technique using ion-induced emission of secondary particles for material analysis is Secondary Ion Mass Spectrometry (SIMS). At the high energy side (tens of MeV) the energy deposition close to the surface is predominately electronic and has been shown to significantly enhance yields of large molecular secondary ions (Jeromel et al., 2014). At the low energy side (up to tens of keV) primary ion beams of comparably heavy projectiles like $Cs^+$, $O_2^+$, $O^+$, $Ar^+$, and $Ga^+$ are upon impact on the surface causing large collision cascades and ejection of secondary particles from near the surface (Werner, 1974). In this regime, the energy transfer process is dominated by nuclear sputtering causing erosion of the matrix material and alteration of the surface structure beneficial for depth profiling (Zhu et al., 2012), magnetron sputtering (Farhadizadeh and Kozák, 2022), surface cleaning (Seah and Nunney, 2010) as well as milling (Vasile et al., 1997). While a considerable effort has been made to understand secondary ion emission in the low keV and MeV energy regime through both experiments and modelling (Fenyö et al., 1990; Taglauer et al., 1980), the regime of sub-MeV ion energy, where often both electronic and nuclear energy deposition contribute significantly, has received considerably less attention. Using a Time-of-Flight Medium Energy Ion Scattering (TOF-MEIS) system, a correlation between the yield of desorbed positively charged surface contaminants and electronic stopping power was observed using H and He ions (Kobayashi et al., 2014; Lohmann and Primetzhofer, 2018). A similar effect was described by Baragiola [22] when measuring the secondary electron yield after impact of 30 keV $H_2$ on surfaces contaminated with hydrocarbons. Thomas et al. (Thomas et al., 1985), studied desorption induced by hydrogen cluster ions with several hundred keV showing additive effects of the cluster components. Recently, when patterning 2D materials with heavy ions of hundreds of keV simultaneous surface cleaning even at non-irradiated areas was observed, due to electronic sputtering in combination with surface diffusion (Tran et al., 2023).

We employ He, B, and Ne ions with several tens to hundreds of keV primary energy to shed light on the relative contribution of electronic and nuclear energy deposition in the ion-induced emission of secondary ions. We employ a Time-of-Flight approach due to the possibility of using extremely low currents in the range of a few fA, proven effective for analysis of sensitive systems without alteration of their composition, surfaces and structure [20]. The sputtering/desorption mechanism is studied simultaneously from two identical surfaces of a 50 nm thick self-supporting Si membrane acquiring information in backscattering and more unconventionally transmission geometry, providing insight on effects of directionality in the momentum transfer.

**Experiment and method**

All experiments were performed using the ToF-MEIS setup at Uppsala University. The ion source in combination with the accelerator from Danfysik produces a wide range of positive ions (Linnarsson et al., 2012; Sortica et al., 2020) with energies in a range from 5 to 350 keV for singly charged ions. Pulsing of the beam with typical pulse durations of 1 – 2 ns is achieved by the electrostatic sweeping of the beam with two orthogonal choppers. The cross-section of the beam on target is about 1 mm$^2$ with the current limited to at most a few pA. The chamber features a 6-axis goniometer holding the sample and a large position-sensitive microchannel plate (MCP) detector from Roentdek ("See http://www.roentdek.com for details on the detectors and software," 2024). The detector covers scattering angles ±11 degrees and can be rotated around the sample at a fixed radius of 290 mm. The measurements were conducted on a 50 nm thick single-crystalline Si (100) membrane purchased from Norcada Inc.("See https://www.norcada.com/products/silicon-membranes," 2024). The sample was transferred from ambient conditions to the chamber without any prior cleaning. To prevent unintentional channelling in the present experiments, the membrane was tilted 7° around the in-plane horizontal axis of the goniometer and 30° around the in-plane vertical axis. The detector was positioned in two configurations: backscattering (BS) with the detector situated at scattering angle 150° from the incoming beam and transmission (TM), with the detector placed at -30° (see Figure 1a), both resulting in parallel sample and detector surfaces. To acquire the projectile impact rate, which is essential for normalization and verification of the beam stability, in-between each measurement the detector was periodically positioned perpendicular to the axis of the incoming beam. The rate of the incoming projectiles was controlled to not exceed 10 000 particles/s ensuring on average a single ion impact per pulse. The measuring intervals for each geometry were 180 s and in total three spectra were acquired on each detection side. The perpendicular position of the detector was further employed to confirm the pseudo-random orientation of the sample assessed based on the recorded blocking pattern (Holeňák et al., 2020). The yield of primary projectiles scattered outside of the detector was reconstructed by the Monte Carlo Binary Collision Approximation code SIIMPL (Holeňák et al., 2024). The definition of the crystal orientation and beam parameters allowed for the simulation of the expected angular distribution of transmitted projectiles as well as the number of projectiles implanted in the membrane. The latter effect accounted for a correction of at most about 3% for heaviest projectiles and lowest energies, way below other experimental uncertainties. The effect of trajectories not ending in the detector on the other hand required a correction of the assumed incident fluence up to 300% for e.g. 40 keV B and was therefore properly implemented in the evaluation.

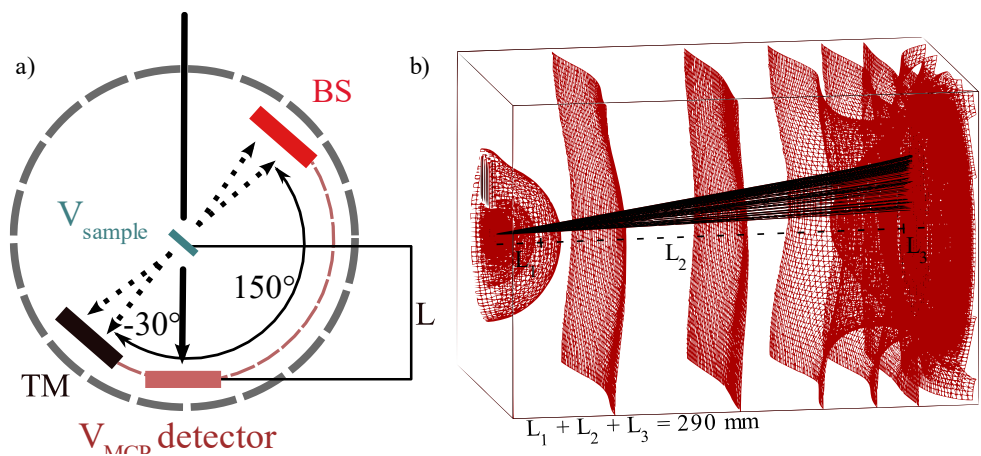

Figure 1. a) Schematics of the experimental geometry. To preserve maximum azimuthal symmetry of the electrostatic field collecting the secondary ions the surfaces of the sample and detector are oriented parallel to each other. b) SIMION simulation of the electrostatic field between the sample holder (left) and the MCP detector (right). Black lines demonstrate the typical trajectory of desorbed low-energy particles.

The collection of all positive secondary ions emitted upon ion impact was ensured by biasing the specially designed sample holder. The sample holder was set to a potential of 700 V, the grid, positioned 1.8 cm from the detector, was grounded, and the front side of the detector was set to -2500 V. (see Figure 1.b) The field between the detector and the sample was simulated with SIMION particle trajectory simulation software (Manura and Dahl, n.d.). Aside from the acceleration of secondary ions, the SIMION simulation confirmed that any particle leaving the centre of the sample within a solid angle of $2\pi$ would be focused on the detector. All particles with equivalent charge are accelerated to the same energy but reach the detector at different times due to different velocities enabling the separation of the particles by mass. The electrostatic field created between MCP and the sample holder qualitatively can be separated in three regions with different characteristic lengths: acceleration region $L_1$, drift region $L_2$, and detection region $L_3$. The measured time of flight of the particle can be converted to its mass as it is given by (Lohmann and Primetzhofer, 2018):

$$t = \frac{\sqrt{2m} \times L_1}{\sqrt{E_0} + \sqrt{E_0 + qV_S}} + \frac{m \times L_2}{\sqrt{2(E_0 + qV_S)}} + \frac{\sqrt{2m} \times L_3}{\sqrt{E_0 + qV_S} + \sqrt{E_0 + q(V_S - V_{MCP})}}$$

where $E_0$ is the initial energy of the particle leaving the sample. $V_S$, and $V_{MCP}$ are applied voltages on the sample and MCP. The final yield of desorbed/sputtered species yield was obtained by normalizing the integrated peak in the mass/charge spectra for BS and TM measurements to the respective corrected number of transmitted projectiles.

**Results**

Spectra of secondary ions emitted upon the impact of 220 keV Ne primary ions in backscattering and transmission geometry, converted from time-of-flight to mass-over-charge (amu/q) are shown in Figure 2. Within the measurement window 0-70 amu/q, multiple peaks assigned to hydrogen and mainly hydrocarbon molecules $C_xH_y$ with an odd number of hydrogen atoms could be clearly identified. The spectrum thus reveals the presence of contaminants on the silicon membrane with a high sensitivity to relatively small amounts of carbon atoms. The expected surface coverage of C atoms in the employed system, previously reported by transmission MEIS-ERDA, is $1.4 \pm 0.4 \; 10^{15}$ atoms/cm$^2$ (Holeňák et al., 2022). Hence, the approach appears promising in being capable of identifying contaminants with a surface coverage down to a few percent. The overall mass resolution is better than 1 amu. The majority of the peaks except one show a narrow symmetric distribution indicating a limited spread of initial energy of particles leaving the sample surface. The peak around amu/q 28 where silicon atoms are expected, exhibits a tail towards lower mass values. The initial energy of the sputtered atoms is expected to feature an inverse square-law like distribution towards higher energies up to keV (Dullni, 1984), in accordance with the experimental observation. When employing He ions, the corresponding peak was found much smaller without the high-energy tail. A distinct increase in the normalized overall yield is observed for TM over BS. This effect was found at all measured energies for both Neon and Boron but was absent in the measurement with Helium.

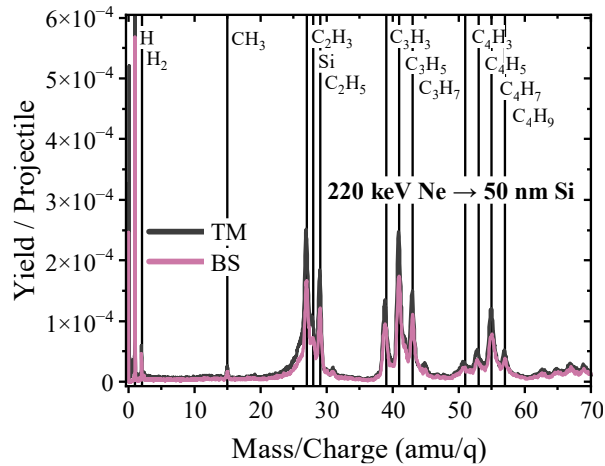

Figure 2. A mass-over-charge spectrum of secondary ions acquired membrane in transmission (TM - black line) and backscattering geometry (BS - purple line) using 220 keV Ne primary ions incident onto a 50 nm thick silicon. Prominent peaks in the spectra are highlighted by vertical lines with chemical formulae of contaminants/hydrocarbon fragments.

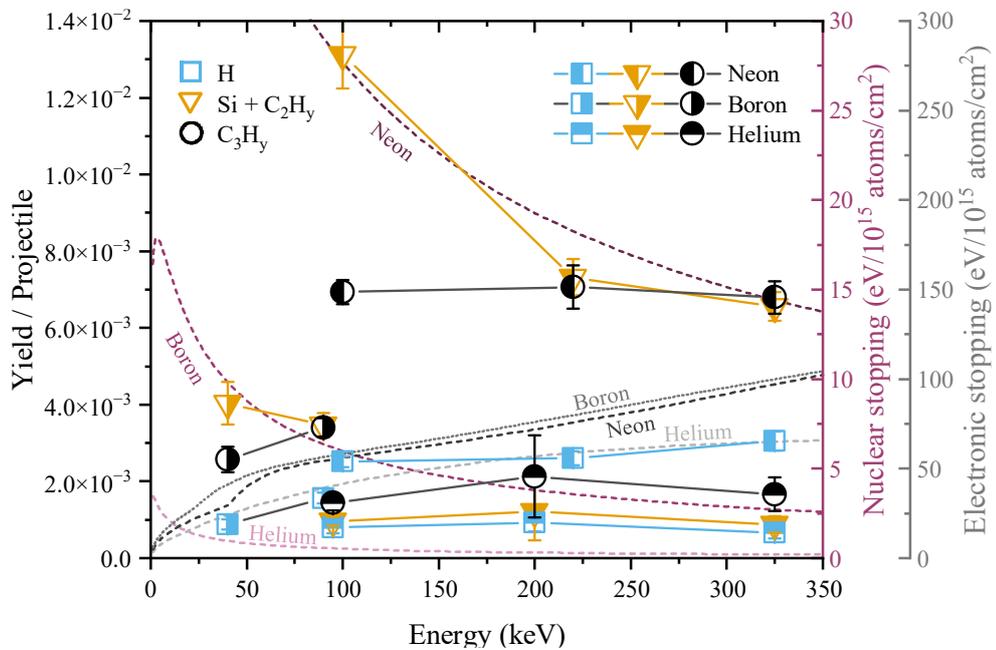

Figure 3. Yields of H, Si + $C_2H_y$, and $C_3H_y$ (left vertical axis) as a function of incident energy for He, B and Ne. The corresponding stopping power values of He, B and Ne in Si (right vertical axes) were obtained from SRIM (Ziegler et al., 2010).

Figure 3 shows a compilation of atomic and molecular secondary ion yields obtained from all employed projectiles and energies. Three groups of signals with different amu/q are presented: (i) atomic hydrogen H, (ii) the integrated signal ranging from 20-30 amu/q including Si and $C_2H_y$ and (iii) presumably $C_3H_y$ molecules within 37-47 amu/q. The figure furthermore features curves showing the electronic and nuclear stopping power for respective projectiles in Si extracted from SRIM. For each of the identified groups, the highest yield is observed for Ne and the lowest for He. All data points for He projectiles show only a weak dependence on the energy of the primary ions. The same is true for H and $C_3H_y$ clusters measured with Ne. For B, H and $C_3H_y$ show an increase aligned with a steeper increase in electronic stopping. The (ii) group including Si displays an apparent scaling with nuclear stopping for Ne and B and, above that, the relative yields of Si for Ne and B agree well with the nuclear stopping ratio.

**Discussion**

The yields of desorbed species for Helium show only a weak energy dependence which makes it difficult to identify either nuclear or electronic energy deposition as the sole root cause for desorption of secondary ions from the surface as previously observed by (Kobayashi et al., 2014) where the focus, however, was on desorption of matrix atoms from an ionic crystal. When comparing the yields of H and $C_3H_y$ for B and He the difference can be ascribed to the considerably higher electronic stopping power for B ions, indicating a prominent role of electronic excitations in the desorption process. This interpretation is solidified also by the energy dependence of the yields for bulk and surface species as observed for B as a projectile. However, the, in contrast, almost twofold increase in yield when comparing data from Ne cannot be of the same character since the electronic stopping power is actually lower for Ne than for B. As the nuclear stopping power for Ne in Si, however, is significantly higher than for

B, this observation suggests a synergetic effect between both nuclear and electronic stopping powers. Related observations where nuclear collisions increase the influence of electronic deposition were found in several systems (Jin et al., 2018; Schenkel et al., 1998; Thomas et al., 1985).

A rather simple explanation would be, that carbon and/or hydrogen nuclei are upon the impact of the projectile recoiled from an intact molecule, which subsequently is desorbed due to the changed molecular configuration leading to ionization. The probability of a collision sufficient to displace a carbon atom from a molecular bond can be crudely estimated using a binary collision formalism. The impact area defined by the minimum impact parameter necessary for an energy transfer of at least 28 eV (energy of a carbon displacement in carbon according to SRIM) is compared to an area occupied by one carbon atom in the layer of known surface density. At projectile energy of 150 keV, the expected yield/projectile of species desorbed by this mechanism would amount to $0.8 \cdot 10^{-3}$, $7.5 \cdot 10^{-3}$ and $25 \cdot 10^{-3}$ for He, B and Ne projectiles respectively. Though not corrected for ionization, the absolute yields exhibit the same magnitude as the measured data. However, unlike the experimental data, the energy scaling of the direct recoil mechanism would show a pronounced increase with decreasing primary ion energy similar to the nuclear stopping curve, which stems from the same nuclear encounter principle. The secondary ion yields of desorbed molecules reported in this work being in the order of $10^{-3}$ for the grouped molecular species in the mass range 1-70 amu/q compares well with the yields reported using keV SIMS (Wong et al., 2003). It has been shown, that detection of larger molecules is often more complicated due to the poor ionization probability $<10^{-3}$ (Wucher, 2018). Energy dependence of the charged fraction could thus also obscure the true charge-integrated yield of specific fragments desorbing from the surface.

As an alternative or synergistic explanation for the observed synergy, it could be reasoned, that the phononic activity induced by nuclear stopping, effectively corresponding to high local temperatures lowers the electronic excitation threshold for electronic desorption. A critical question in this context is, at which time-scale the respective excitations occur at the location of the molecule as well as when the actual desorption takes place, relative to the ion impact. Here, the directionality of the collision cascade may have an effect on the relative timing. Despite numerous efforts (Golombek et al., 2021; Kalkhoff et al., 2023; Mihaila et al., 2023; Ntemou et al., n.d.), at present, experiments accessing the corresponding time domains are not feasible.

Employing the transmission geometry further corroborates the important role of the nuclear stopping power when it approaches a similar order of magnitude to the electronic stopping power. Considerably higher yields are found in TM for B and Ne but not for He. As shown in Figure 4, the yield of Si sputtered by 220 keV Ne on the TM side of the membrane increases by around 53 %. The yields of H and $C_3H_y$ molecules display a slightly lower scaling (about 26% and 36% respectively). Nevertheless, the expected increase of the nuclear stopping power acting on a Ne projectile exiting the membrane on the TM side accounts for only about 12%. Two plausible explanations for the excessive increase in the desorption yield on the transmission side can be offered: At first, we can consider the inherit directionality of the collision cascade prompted by the projectile the momentum of which points towards the exit surface (Hou and Eckstein, 1990; Schlueter et al., 2020). In consequence for recoiled Si atoms, kinematics permits them to exit the membrane in a wider angular range on the TM side (Holeňák et al., 2022). This will trivially increase the yield of Si but is also expected to enhance ion-induced as well as potentially additional recoil-induced desorption of surface contaminants. Secondly, the change in the charge state distribution of the exiting projectiles featuring a non-negligible fraction of projectiles with higher charge state leading to higher electronic stopping power (Holeňák et al., 2021; Juaristi et al., 1999). These effects are likely to act simultaneously, rendering the synergy mechanism more complex.

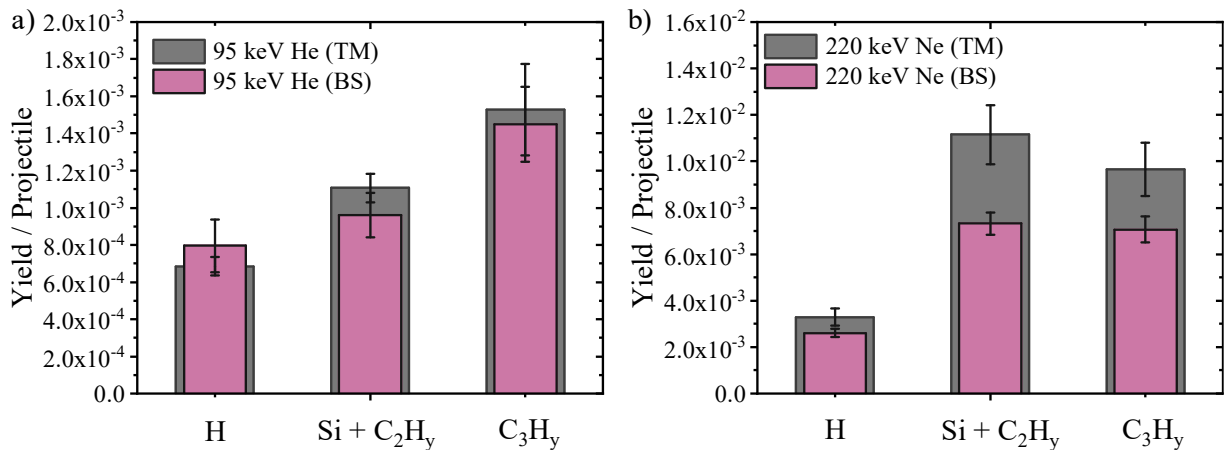

Figure 4. Comparison between backscattering and transmission geometry for a) 95 keV He ions and b) 220 keV Ne ions. A significant, but non-uniform increase in the desorbed yields of the three selected mass groups is found only for Ne.

For a more complete understanding, also charge-integrated measurements of desorbed species would be highly beneficial, however, are not possible in the present set-up. Such an approach would provide the most complete insight, as precise quantitative data regarding the ionization probabilities of secondary particles (Wucher, 2018) are scarce and are further subject to surface chemistry. Transmission experiments with higher initial charge state of the incoming projectile could potentially address the influence of charge-exchange processes on surface ionization and consequent desorption probability discussed in (Baragiola, 1984). Employing again simultaneous recording in transmission geometry would be highly beneficial as the exiting charge state distribution on the transmission side should remain unchanged (Holeňák et al., 2021).

**Summary and outlook**

We studied the emission of secondary ions emitted from both surfaces of a 50 nm thick self-supporting silicon membranes upon transmission of pulsed beams of He, B, and Ne ions with primary energies of 40-325 keV. A clear relation was found between the yield of sputtered Si and the nuclear stopping. The desorption of H and molecular species was shown to be governed by a more complex mechanism. The relative scaling among the employed projectiles revealed a non-trivial synergetic effect between electronic and nuclear energy deposition processes. This effect was further corroborated by the increase in yields observed for different species in transmission. Future experiments with different incident charge states and/or molecular ion beams, as well as a wider range of geometries are considered to be capable of obtaining a better insight into the mechanisms of bond-breaking for surface adsorbents. The findings of this work furthermore enhance the analytical potential of the method applicable in e.g. *in-situ* fabrication and modification of ultrathin layers and in studying catalytic reactions at surfaces.

## Acknowledgements

Accelerator operation was supported by the Swedish Research Council VR-RFI (Contract No. 2019-00191)